# Effect of crystallographic texture on dealloying kinetics and composition of nanoporous gold surface

Ezgi Hatipoğlu[1*], Ayman A. El-Zoka[1,2], Yujun Zhao[1], Stanislav Mráz[3], Jochen M. Schneider[3], Baptiste Gault[1,2], Aparna Saksena[1*]

[1] Max-Planck-Institute for Sustainable Materials, Max-Planck-Straße 1, Düsseldorf 40237, Germany

[2] Departments of Materials, Imperial College London, Royal School of Mines, Prince Consort Rd, South Kensington, London SW7 2AZ, UK

[3] Materials Chemistry, RWTH Aachen University, Kopernikusstrasse. 10, Aachen 52074, Germany

*corresponding authors: e.hatipoglu@mpi-susmat.de, a.saksena@mpi-susmat.de

## Abstract

Nanoporous metals allow for tailoring composition and surface-to-volume ratio, both aspects critical for applications in catalysis. Here, $Ag_{70}Au_{30}$ (±2 at. %) films with a face-centered cubic structure were deposited at 400°C, either {111}-textured or randomly oriented. Upon chemical dealloying, atom probe tomography of the nanoporous structure reveals that the textured film retains a up to 2.7 times higher Ag concentration within the ligaments compared to the randomly oriented film that exhibits ligament coarsening, indicating faster dealloying kinetics. Our study highlights the potential of microstructure engineering in tailoring the properties of nanoporous metals for possible future catalytic and electrochemical applications.

## Keywords:
Nanoporous Gold, Thin Films, Atom Probe Tomography, Texture, Dealloying

Dealloying has become an important method for obtaining highly functional nanoporous materials. By the selective dissolution of the less noble element from a solid solution in acidic environments, an open-pore, bicontinuous structure of the more noble element can be achieved [1]. The Ag-Au system has been widely investigated for nanoporous gold (NPG) formation [2,3], showing promising performance as catalysts for e.g. methanol oxidation in fuel cells [4]. It has been demonstrated that Ag as an impurity is responsible for the high catalytic activity of Au powders, foam, and nanotubes. Compared to pure Au or pure Ag, Au-Ag alloy catalysts exhibit higher activity towards CO or $H_2$ oxidation [5–7]. Therefore, gaining control over the Ag concentration in the ligaments can tailor the catalytic properties of NPG.

The microstructure of the precursor material, for instance, its grain size, texture, etc., is critical for understanding nanoporosity formation and the composition evolution of the ligaments. Grain boundaries can serve as fast diffusion pathways or locally slow down the dissolution of the less noble element [8–10]. Badwe et al. investigated the intergranular cracking in Ag-Au alloys by closely examining the role of grain boundaries in guiding the evolution of the dealloyed layer and its interface with the parent phase, acting as a fast diffusion pathway for dissolution [9,11]. Gholamzadeh et al. demonstrated that the crystallographic orientation of Alloy 800 significantly influences the thickness and size of the ligament formation of nanoporous Ni-rich film, showing grains oriented near the {111} plane tend to form finer ligaments compared to other grain orientations [8]. Although these findings successfully demonstrate the role of the microstructure on the evolution of ligament morphology during dealloying, it remains unclear how the microstructural features affect the local composition of these ligaments.

To deconvolute and reveal such microstructure-dependent mechanisms, using thin films instead of bulk solid solutions can be beneficial. Using physical vapor deposition, precise control over thin film thickness and composition [12], and microstructural defect density can be achieved [13], offering an opportunity to control the kinetics of the nanoporous structure formation [14]. Au-Ag thin films have been investigated, where the influence of synthesis routes [10], precursor alloy composition [15], and adhesion layer [14] have been optimized to achieve a functional crack-free nanoporous structure.

Here, we investigated two different samples deposited via magnetron sputtering under identical conditions but modified the reactive adhesion/ inter-layer, which led to a film where the grain orientation was primarily {111}-textured, while for the other, grains did not show a preferred orientation and were randomly oriented. The influence of grain orientation on chemical dealloying and ligament composition was analyzed at the nanoscale to advance the understanding of the dealloying mechanisms and gain deeper insights into the microstructure's role in nanoporous structure formation.

$Ag_{70}Au_{30}$ thin films were deposited using direct current magnetron sputtering in high-purity Ar (99.999% purity) at a pressure of 0.4 Pa, where a base pressure of ≤ 7 x $10^{-5}$ Pa was maintained before every deposition, using the setup described in reference [16]. For both samples, the $Ag_{70}Au_{30}$ film was synthesized by co-sputtering Ag (99.99% purity) and Au (99.9% purity) targets at 400°C by applying power densities of 6.1 W/cm$^2$ and 2 W/cm$^2$, respectively. The texture variation between the samples is attained by modifying the reactive interlayer, achieved by deposition of Ti on a Si (001) substrate, with a power density of 2.5 W/cm$^2$ where the deposition time is varied, 10 or 20 minutes. They are termed as $Ag_{70}Au_{30}$-T (textured) and $Ag_{70}Au_{30}$-NT (non-textured), respectively. Discussion pertaining to the reactive interlayer can be found in the supplementary information (see Fig. S1 and Fig. S2). The substrates were continuously rotated during deposition to ensure a uniform composition and film thickness across the sample's surface.

Microstructural characterization was performed using a Zeiss Sigma scanning electron microscope (SEM, Carl Zeiss SMT, AG, Germany) equipped with an EDAX AMETEK detector. Backscatter electron (BSE) imaging was performed using a Zeiss Merlin scanning electron microscope (SEM) equipped with a Gemini-type field emission gun. Electron backscattered diffraction (EBSD) was obtained with a step size of 0.05 µm, and the grain orientation and misorientation analysis were carried out using the EDAX OIM Analysis 8 software. Energy dispersive X-ray spectroscopy (EDS) was performed with an accelerating voltage of 15 kV and a probe current of 2–5 nA. To investigate the Ti adhesion/ reaction inter-layers, the transmission electron microscopy (TEM) lamellae were prepared by site-specific lift-out and polishing procedure [17] using a dual-beam focused ion beam (FIB) (Helios, Nanolab 600i, FEI). The scanning transmission electron microscopy was conducted with a JEOL JEM-2100 PLUS.

Needle-shaped atom probe tomography (APT) specimens were prepared using the FIB (Helios, Nanolab 600, FEI) by following the protocol in reference [18]. The local composition, both before and after dealloying, was characterized using APT on a local electrode atom probe LEAP 5000 XR (Cameca Instrument Inc., Ametek, Madison, USA), in laser-pulsing mode, with 60-70 pJ laser pulse energy at a repetition rate of 100 kHz, and with an average detection rate of 5 ions per 1000 pulses at 60 K during the measurement. The collected data were reconstructed and analyzed using AP Suite 6.3.

Small sections (~4×10 mm²) of the as-deposited thin films were cut and chemically dealloyed by keeping them in concentrated nitric acid for 30 minutes at room temperature (RT). Afterward, the samples were washed with ultra-pure (0.055 µS/cm) deionized water (DI water) to remove residual acid and prevent further dealloying. NPG formed from $Ag_{70}Au_{30}$-T referred to as NPG-T, while NPG formed from $Ag_{70}Au_{30}$-NT is referred to as NPG-NT.

Samples were kept in DI water overnight before Cu electrodeposition was performed to fill the nanopores and facilitate specimen preparation for APT analysis, following the procedure detailed in Ref. [19]. The sample was immersed in a 1 M $CuSO_4$ + 0.1 M $H_2SO_4$ solution by using a three-electrode system, in which a mercury sulfate electrode (MSE) was used as the reference electrode, and a platinum wire was used as the counter electrode. Cyclic voltammetry was conducted from the open-circuit potential to -0.4 V vs MSE with a scan rate of 0.5 mV/s, with three scans to determine the optimal deposition potential (Fig. S3). Chronoamperometry was then performed in the same solution for 30 minutes at - 400 mV vs MSE to deposit Cu onto the porous layer (Inset of Fig. S3). These electrochemical experiments were conducted at RT.

Fig. 1 presents the SEM and EBSD characterization of as-deposited $Ag_{70}Au_{30}$ thin films. Fig. 1 a-b are BSE images revealing the surface morphology of the $Ag_{70}Au_{30}$-T and $Ag_{70}Au_{30}$-NT, respectively. The grain size distribution was analyzed using the line intercept method following ASTM E112 [20], based on measurements from BSE images. Grain boundary intercepts were counted manually, and the average grain size for both $Ag_{70}Au_{30}$-T and $Ag_{70}Au_{30}$-NT showed a similar trend (1.21 µm ± 0.25 µm and 1.23 µm ± 0.26 µm, respectively). Fig. 1c is the corresponding EBSD map, showing inverse pole figure (IPF) with image quality (IQ), demonstrating that the microstructure on $Ag_{70}Au_{30}$-T exhibits a strong {111} texture along the normal direction, suggesting preferred grain orientation during growth. Meanwhile, Fig. 1d reveals that $Ag_{70}Au_{30}$-NT has a random grain orientation, without any significant preferred texture. Different grain boundary types and ratios are discussed in Table S1. Grain morphology for both samples is equiaxed as demonstrated by the back-scattered electron cross-section images acquired after FIB cuts performed at a 30° angle from the surface (Fig. S4).

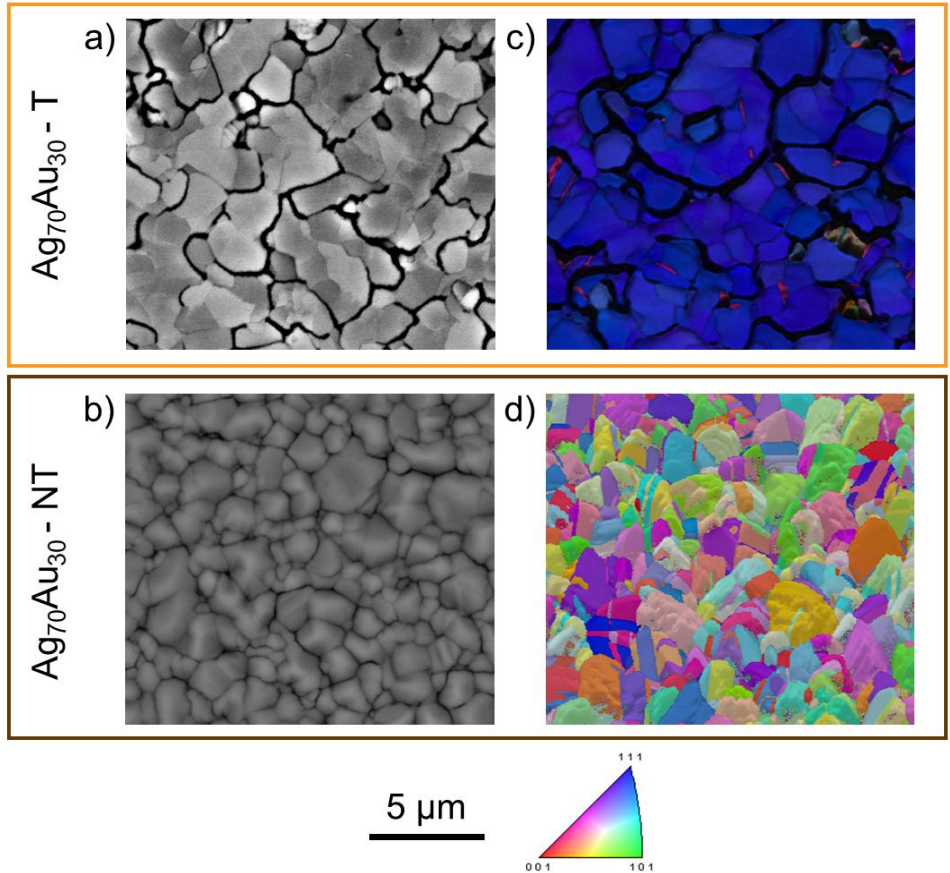

Fig. 1. Backscattered electron (BSE) images of a) $Ag_{70}Au_{30}$-T (Textured), and b) $Ag_{70}Au_{30}$-NT (Non-Textured), Electron Back-Scattered Diffraction (EBSD) image quality (IQ) map and inverse pole figure (IPF) map along the normal direction of c) $Ag_{70}Au_{30}$-T, and d) $Ag_{70}Au_{30}$-NT.

The elemental distribution of Ag and Au in the thin films analyzed by APT is reported in Fig. 2 from the surface to the bulk. Both samples exhibit a homogeneous distribution of Ag and Au, with a bulk composition of $Ag_{70}Au_{30}$ (±2 at. %) (Fig. 2a-b). Based on these results and the one-dimensional (1D) concentration profile from the region of interest (ROI) (Fig. 2c-d), no significant local compositional variations are observed at the atomic scale that could influence the dealloying behavior of either of the two samples.

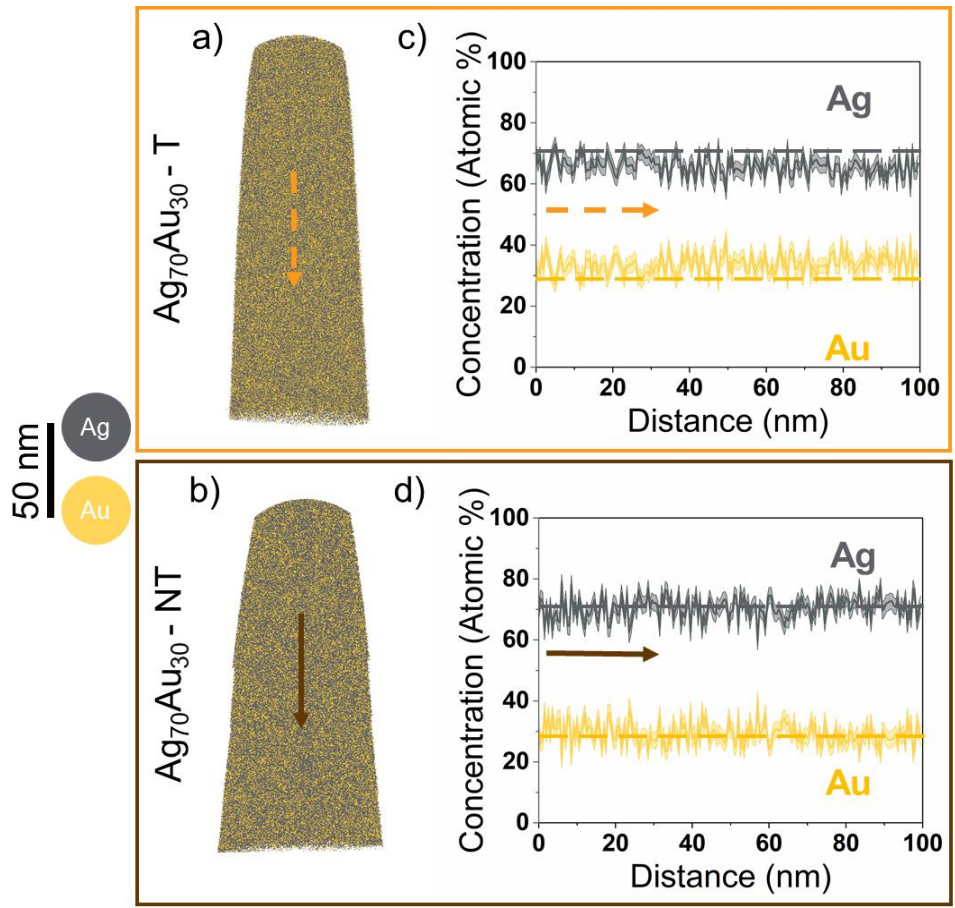

*Fig. 2. The reconstructed datasets obtained via Atom Probe Tomography (APT) from a) $Ag_{70}Au_{30}$-T (Textured), and b) $Ag_{70}Au_{30}$-NT (Non-Textured), the one-dimensional profile across the region highlighted in (a) and (b) c) $Ag_{70}Au_{30}$-T, and d) $Ag_{70}Au_{30}$-NT, respectively.*

Fig. 3a-b include secondary electron (SE) images of NPG-T and NPG-NT, respectively, clearly showing the nanoporosity formation in $Ag_{70}Au_{30}$-T thin films. According to the EDS analysis from the same area, NPG-T has 18.6 at. % Ag (± 3 at. %) and 81.4 at. % Au (± 2 at. %) while NPG-NT has 7.0at. % Ag (± 2 at. %) and 93.0 at. % Au (± 3 at. %). These results demonstrate that, despite undergoing the same dealloying period, the residual Ag content in NPG-T and NPG-NT thin films differs significantly by nearly 2.7 times.

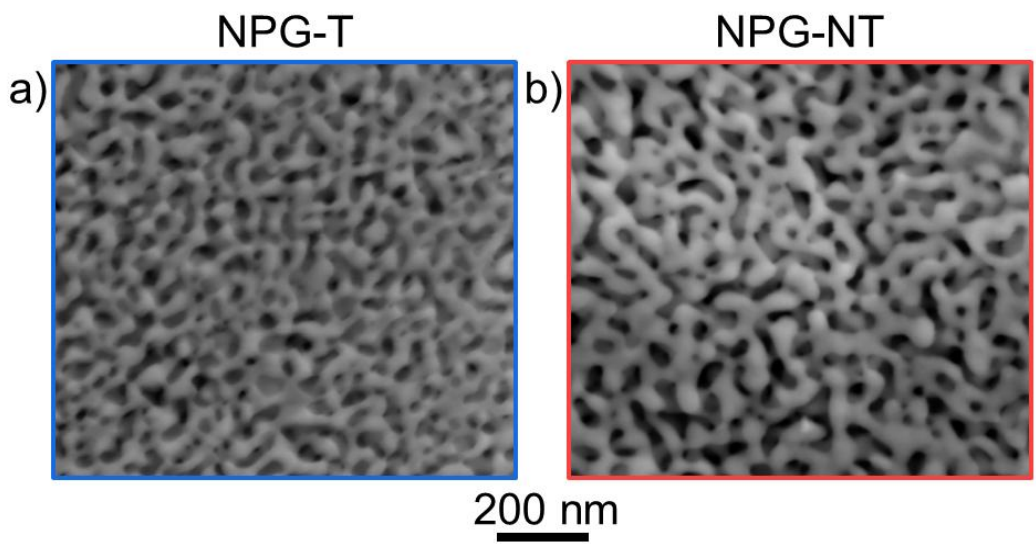

*Fig. 3. Secondary Electron (SE) images of a) Nanoporous Gold – Textured (NPG-T), and b) Nanoporous Gold – Non-Textured (NPG-NT).*

To further investigate the effect of different dealloying kinetics on these two NPG thin films, Cu was electrodeposited to fill the pores and enable APT [19], as schematically depicted in Fig. 4a. Fig. 4b-c are side views of the reconstructed APT datasets, respectively, for the Cu-filled NPG formed from the NPG-T and NPG-NT. The reconstructed copper ions are plotted as orange dots, enabling clear visualization of the porosity. The ligaments contain gold, plotted in yellow, and silver in dark gray. The bulk compositions obtained from APT indicate that the NPG-T sample exhibits a lower Au: Ag ratio (~5) compared to the NPG-NT sample (~12). These findings follow the information obtained via EDS; despite undergoing the same dealloying conditions, the lack of a preferred crystallographic texture appears to make $Ag_{70}Au_{30}$-NT more prone to dealloying than $Ag_{70}Au_{30}$-T.

In the NPG-T sample, the composition profile extracted from the APT data and plotted in Fig. 4d reveals that Ag is predominantly located in the core of the ligaments, with an approximate concentration of 40%, indicating that the sample is still in the primary dealloying stage [21]. In contrast, due to the faster dealloying kinetics in NPG-NT, some ligaments have already coarsened, leading to the dissolution of Ag from the core of ligaments as indicated by the 1D composition profile in Fig. 4e. Ag diffusion from the core toward the surface of ligaments in NPG-T was observed in Fig. 4f, suggesting that Ag continues to be dissolved. NPG-NT exhibited no detectable Ag at the ligament surface, as shown by the compositional profile (Fig. 4g) along the ROI in NPG-NT. This suggests that the Ag on the surface of the ligaments has already dissolved, and the overall Ag content within the ligaments is significantly lower compared to NPG-T. These findings highlight the significant changes in the compositions of ligaments for both NPG thin films.

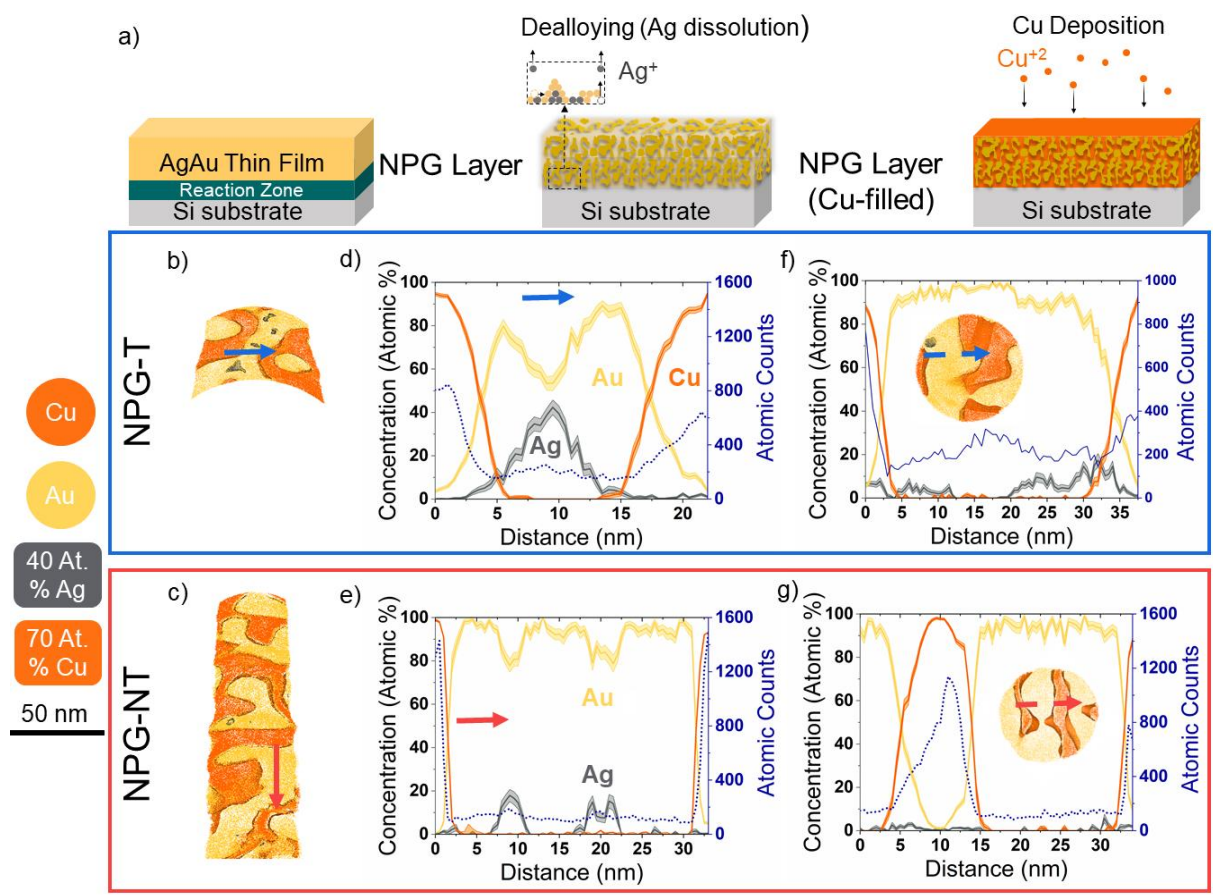

*Fig. 4. (a) Schematic illustration of experimental workflow for dealloying and copper filling, adapted from [22], side views of reconstructed Atom Probe Tomography (APT) datasets b) Nanoporous Gold – Textured (NPG-T), and c) Nanoporous Gold- Non-textured (NPG-NT) showing iso-concentration surfaces of 40 at. % Ag and 70 at. % Cu, one-dimensional (1D) profiles across the regions highlighted in (b) and (c) ((d) and (e), respectively), 1D profiles of top views across the regions (f) and (g) highlighted in the inset of graphs.*

Although the electrodeposition of Cu effectively creates a continuous material enabling specimen preparation and APT analysis of NPG, it results in a heterogeneous structure composed of two different phases, namely Cu and NPG. This heterogeneity can lead to some challenges during APT data reconstruction due to variations of evaporation fields in different phases [23]. Although the reconstruction protocol assumed a hemispherical end shape for the specimen [24], these differences in evaporation fields resulted in changes in the specimen's surface shape. A lower local radius of curvature develops in the regions containing the phase with the higher evaporation field, while a larger radius develops in low-evaporation-field regions. These local deviations in curvature cause aberrations in the ion trajectories, and potentially lead to compositional overlap between phases, in this case between porous structure (copper) and ligaments (gold) [23].

These differences in evaporation fields could lead to a different detection probability of one element over the others in the material [25]. Typically, the element affected is the one with the lowest evaporation field. Atoms of this element field evaporate in between pulses, at the electrostatic field, precluding the elemental identification of the ions via their time-of-flight, and these ions contribute to the background in the mass spectrum. Since this is dependent on the electrostatic field, we used the post-ionization model of Kingham [26] to estimate its intensity based on the ratio of the charge states of a specific element during the APT measurement. Here, we chose the two detected charge states of Au ($Au^+$ and $Au^{+2}$) and compared them with the Ag/Au ratio along the length of the analyzed data for both as-deposited and dealloyed $Ag_{70}Au_{30}$-T and $Ag_{70}Au_{30}$-NT samples, Fig. 5. According to this model [26], the estimated evaporation fields only change slightly between ~27 V/nm and ~31 V/nm across the different measurements. No clear and discernible trend appears in the data that could be indicative of a systematic loss of Ag. This guarantees that the significant changes in Ag composition in the ligaments are intrinsic to the material itself, not related to measurement artefacts.

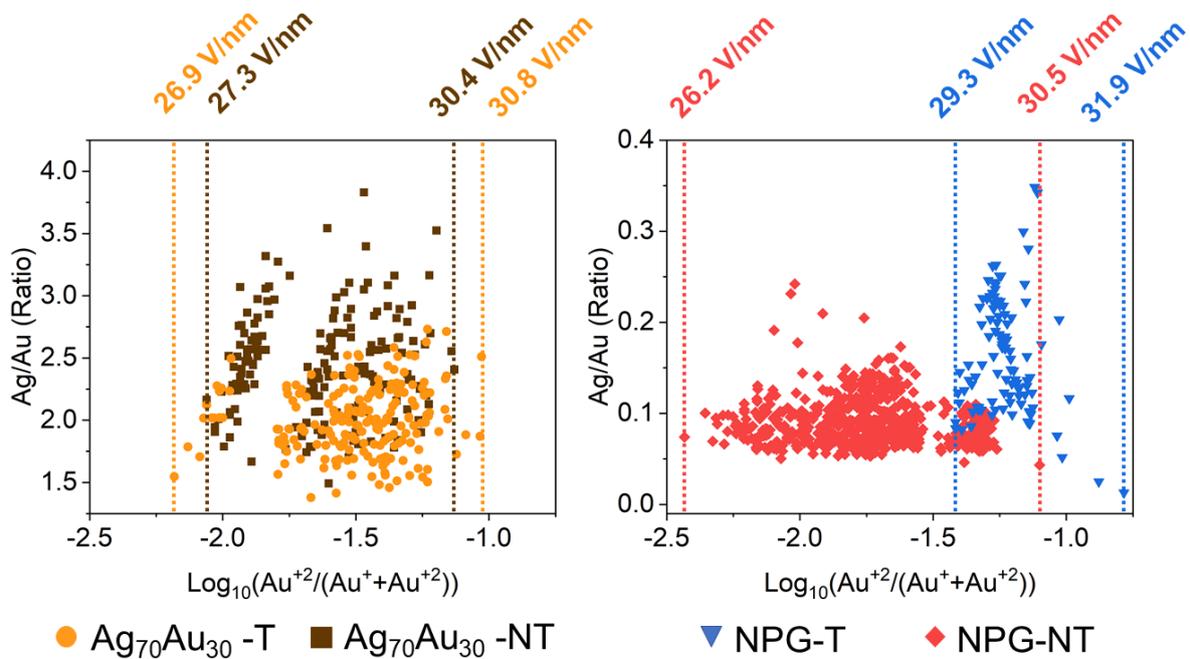

*Fig. 5. The relative abundance of doubly charged gold ions, plotted logarithmically for all APT specimens analyzed for as-deposited (left) and dealloyed (right).*

In combination, based on the SEM, EBSD, and APT, we can conclude that the kinetics of dealloying are faster in the $Ag_{70}Au_{30}$-NT sample compared to the $Ag_{70}Au_{30}$-T, leading to a lower content of Ag dissolved and narrower ligaments for $Ag_{70}Au_{30}$-T. The formation of the porous ligaments can be attributed to a two-step process, which is first, dissolution of the less noble metal, and second, surface diffusion of the more noble metal, consistently exposing the less noble metal to further dissolution [27]. According to DFT studies on the calculation of the activation energies for surface diffusion, it is clear that surface diffusion on the (111) surface (the energy barrier ($E_A$) is 0.22 ± 0.03 eV) is faster than on less packed planes ($E_A$ for the (100) surface is 0.62 ± 0.04 eV) [28]. Therefore, the surface diffusion of Au on the (111) textured samples cannot be the reason for slower dealloying kinetics. A study from Ghlamzadeh et al. had shown the effect of crystallographic orientation in the dealloying of Alloy 800 (Fe-Ni-Cr), also a face-centered-cubic alloy. The higher resistance to dealloying is ascribed to slower dissolution kinetics of the less noble metal from the close-packed {111} planes [8]. A slower kinetics of Ag dissolution from the as-deposited $Ag_{70}Au_{30}$-T thin films is consistent with this observation. While the grain sizes in $Ag_{70}Au_{30}$-T are similar to $Ag_{70}Au_{30}$-NT, and the initial atomic distribution of elements is comparable, the faster dealloying kinetics observed for the $Ag_{70}Au_{30}$-NT films may also be affected by a higher ratio of random high-angle grain boundaries (Table S1) rendering them more chemically active due to their higher energy state and disordered atomic arrangements while $Ag_{70}Au_{30}$-T films have a higher ratio of low-angle grain boundaries by being more resistant [29].

In conclusion, APT helped reveal significant variations in ligament size and composition between two different NPG thin films formed from highly {111}-textured and randomly oriented $Ag_{70}Au_{30}$ thin films. These compositional differences, despite identical dealloying conditions, are attributed to the slower Ag dissolution kinetics along the closed-packed {111} planes that make most of the samples' surface. This finding highlights an important aspect: that the texture of the as-deposited film can be leveraged to tailor ligament composition and morphology. This could help to further investigate and provide deeper insights into dealloying mechanisms at the nanoscale.

## Acknowledgments

The authors thank Uwe Tezins, Christian Bross, and Andreas Sturm for their support at the FIB and APT facilities at MPISusMat. The authors would like to thank Katja Angenendt and Stefan Zaefferer for their support at the SEM facilities at MPI-SusMat.

The authors acknowledge Volker Kree for his support with the TEM infrastructure at MPI-SusMat. EH is grateful for the financial support from the International Max Planck Research School for Sustainable Metallurgy (IMPRS-SusMet). A.A.Z would like to acknowledge support from the Materials Department, Imperial College London, and EPSRC grant EP/V007661/1. Additionally, EH would like to thank Varatharaja Nallathambi for his assistance during sample preparation, Beste Payam, and Taner Özdal for helpful discussions.